# Symmetry Breaking in Haloscope Microwave Cavities

I. Stern, N. Sullivan and D. Tanner

**Abstract** Axion haloscope detectors use microwave cavities permeated by a magnetic field to resonate photons that are converted from axions due to the inverse Primakoff effect. The sensitivity of a detector is proportional to the form factor of the cavity's search mode. Transverse symmetry breaking is used to tune the search modes for scanning across a range of axion masses. However, numerical analysis shows transverse and longitudinal symmetry breaking reduce the sensitivity of the search mode. Simulations also show longitudinal symmetry breaking leads to other undesired consequences like mode mixing and mode crowding. The results complicate axion dark matter searches and further reduce the search capabilities of detectors. The findings of a numerical analysis of symmetry breaking in haloscope microwave cavities are presented.

## 1 Background

The axion particle, first theorized as a solution to the charge conjugation and parity symmetry problem of quantum chromodynamics,[1-3] has been established as a prominent cold dark matter (CDM) candidate.[4] The most sensitive axion search technique is the haloscope detector,[5,6] proposed by Sikivie.[7] The haloscope detector uses a microwave cavity permeated by a strong magnetic field to convert axions to photons via the inverse Primakoff effect.[8] The scan mode of the cavity is tuned across a frequency range to search for CDM axions.

The power measured in the cavity for a specific resonant mode is given by[9]

$$P_{mnp} \approx g_{a\gamma\gamma}^2 \frac{\rho_a}{m_a} B_0^2 V C_{mnp} Q_L, \qquad (1)$$

where the indices $m$, $n$, and $p$ identify the mode. The axion parameters are given by $g_{a\gamma\gamma}$, the axion-photon coupling constant, $m_a$, the mass of the axion, and $\rho_a$, the local mass density. The parameters of the detector are given by $B_0$, the magnetic field strength, $V$, the volume of the cavity, and $Q_L$, the loaded quality factor of the cavity (assumed to be less than the kinetic energy spread of the axion at the Earth). $C_{mnp}$ is the normalized form factor describing the coupling of the axion conversion to a specific cavity mode, derived from the Lagrangian for the axion-photon interaction. It is given by

$$C_{mnp} \equiv \frac{\left|\int d^3x \, \mathbf{B_0} \cdot \mathbf{E_{mnp}(x)}\right|^2}{B_0^2 V \int d^3x \, \varepsilon(\mathbf{x}) |\mathbf{E_{mnp}(x)}|^2}, \tag{2}$$

where $\mathbf{E_{mnp}}$ is the electric field of the mode an $\varepsilon$ is the permittivity within the cavity normalized to vacuum.

## 2 Microwave Cavity Theory

When the permeability in a cavity is homogenous, the modes must satisfy[10]

$$\left(\nabla^2 + 2\pi\mu\varepsilon f^2\right)\begin{Bmatrix}\mathbf{E}\\\mathbf{B}\end{Bmatrix} = \begin{Bmatrix}\nabla(\nabla \cdot \mathbf{E})\\(\nabla \times \mathbf{B}) \times \frac{\nabla\varepsilon}{\varepsilon}\end{Bmatrix}, \tag{3}$$

where $f$ is the mode frequency, and $\mu$ and $\varepsilon$ are the permeability and permittivity within the cavity, respectively. If the permittivity inside the cavity is also constant, Eq. 3 reduces to the eigenvalue problem[11]

$$\left(\nabla_t^2 + 2\pi\mu\varepsilon f^2\right)\begin{Bmatrix}\mathbf{E}\\\mathbf{B}\end{Bmatrix} = 0. \tag{4}$$

For cylindrical cavities, the modes are standing waves consisting of an oscillating field with a constant cross-section, traversing along the longitudinal ($z$) axis. The result is three sets of infinite orthogonal modes, transvers electric (TE), transverse magnetic (TM), and transverse electromagnetic (TEM). The orthogonality applies to the field types independently ($\mathbf{E_i} \cdot \mathbf{E_j}$ and $\mathbf{B_i} \cdot \mathbf{B_j}$) as well as combined ($\mathbf{E_i} \cdot \mathbf{B_j}$).

If the permittivity inside the cavity is not constant along the $z$-axis, the modes are standing waves with varying cross-section, breaking longitudinal symmetry. Specifically, when the permittivity has discontinuities, such as internal boundaries, the orthogonality of the field types independently is broken.[12] The orthogonality breaking gives modes that are not pure TM, TE, or TEM, but some mixed or localized mode. These modes can have significant effects on the sensitivity and search capabilities of a haloscope detector.

Equation 2 shows that the sensitivity of a haloscope detector is strongly dependant on the relationship between the static magnetic field and the orientation of the electric field of the search mode. Typically, a haloscope detector uses a circular-cylinder shaped microwave cavity and a solenoidal superconducting magnet. In this configuration, the search mode would need sufficient electric field pointed in the $z$-axis to observe the axion-to-photon conversion.

In a circular-cylinder microwave cavity with constant permittivity, only the $TM_{0n0}$ modes would couple to the axion, with the $TM_{010}$ mode having the strongest coupling ($C$). However, because the cavity in an axion haloscope must be tuned to different frequencies to conduct a search, discontinuities in permittivity are almost inevitable. To date, haloscopes are tuned using one or more conducting or dielectric rods that run the length of the cavity in the *z*-direction.[13] The rods are moved transversely to change the frequency of the $TM_{0n0}$ modes. Because the rods are moved, a physical gap must exist between the rod-ends and the endcaps of the cavity. The gaps create a discontinuity in the permittivity in the longitudinal direction, breaking mode orthogonality of the $TM_{0n0}$ modes. This phenomenon has been identified as a capacitance effect.[14]

Therefore, search modes in a haloscope detector are perturbations of a $TM_{0n0}$ mode and frequently a mixed mode. The mixing of TM with TE and TEM modes causes gaps in the frequency scan range of a haloscope. Further, additional modes are formed due to the symmetry breaking, increasing the number of modes at similar frequency to the search mode. This mode crowding increases the difficultly of tracking a search mode, making axion detection more challenging.

## 3 Numerical Analysis

In an effort to quantify the effects of symmetry breaking on search capabilities of a haloscope detector, numerical analysis was conducted to evaluate the impact on form factor (*C*), frequency scan range, and mode crowding. The simulations were conducted with a commercially available three-dimensional finite element modeling program (COMSOL version 5.1). All cavity models had a diameter of 5.375 in. and a height of 10.75 in., and used a single tuning rod of diameter 1.430 in. The maximum mesh size was no more than 1/7$^{th}$ the wavelength and an eigenfrequency solver was used to compute the modes.

Simulations of transverse symmetry breaking showed that the form factor is decreased when symmetry is broken. The result matches similar findings in other numerical analyses.[13,15] More significantly, the simulations showed that transverse symmetry breaking did not break mode orthogonality, and did not result in mode mixing or an increase in mode crowding about the search mode frequency.

Simulations of longitudinal symmetry breaking yielded more valuable information. Adding a gap between the tuning rod and endcaps showed mode orthogonality breaking, which caused significant mode mixing at the point where the search mode had a similar frequency to another mode (i.e., a mode crossing). The mixing caused a gap in the frequency scan range of the search mode. The mixing increased as the gap grew, resulting in a larger gap in the frequency scan range.

Figure 1a-d shows the frequency of the lowest search mode and the corresponding form factor as the tuning rod is moved from the center of the cavity for gap-to-height ratios (*g/L*) of 0, 0.001, 0.003, and 0.005, respectively. The distance from the center of the tuning rod to the center of the cavity, *x*, is shown on the *x*-axis, and normalized by the radius of the cavity, *R*.

The plots demonstrate the effects of a mode crossing and shows a degenerate TE mode and the next higher TM mode ($C \approx 0$). The symmetry breaking causes the TE modes to break degeneracy. When the search mode is tuned to a frequency nearly the same as the TE modes, the search mode mixes with one of the TE modes, forming two mixed modes. The other TE mode does not mix with the search mode and maintains orthogonality as well as a constant frequency.

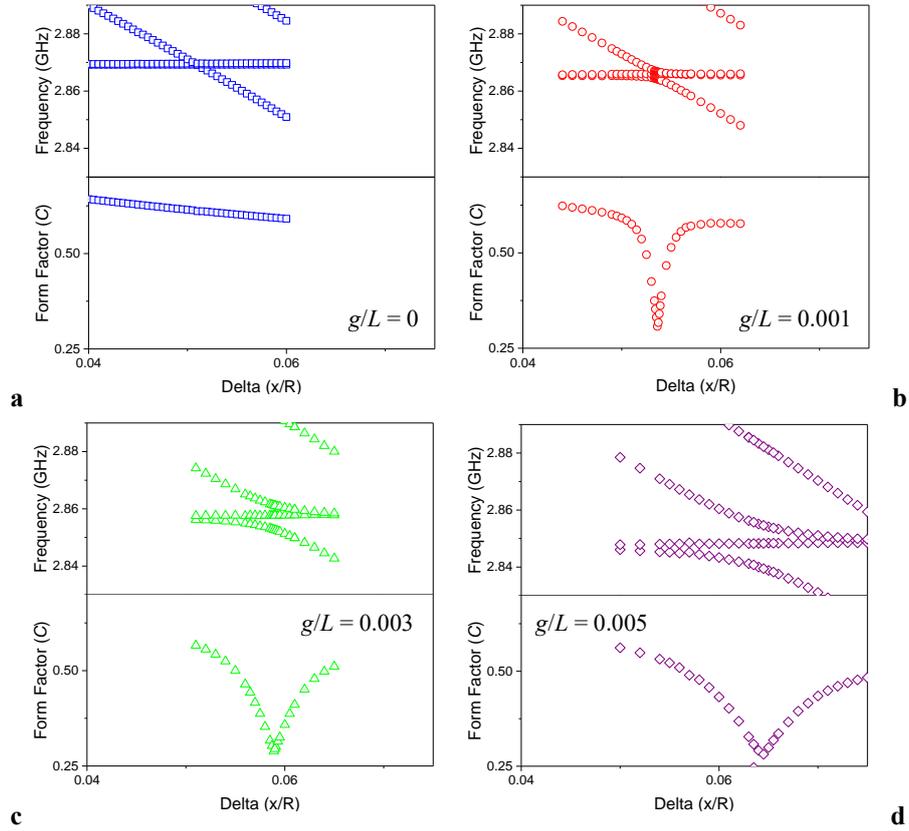

**Fig. 1** Frequency of the lowest search mode and the corresponding form factor as the tuning rod is moved through a mode crossing of the cavity for a rod-end gap of **a)** $g/L = 0$; **b)** $g/L = 0.001$; **c)** $g/L = 0.003$; **d)** $g/L = 0.005$. The distance from the center of the tuning rod to the center of the cavity, $x/R$, is shown on the $x$-axis. As the gap increases, greater mode mixing is observed and the dip in form factor broadens. The next higher TM (non-searchable) mode is also shown.

Mode mixing is depicted as a reduction in form factor and a frequency separation in modes at the crossing. The non-mixing TE mode is depicted as the straight-line data points between the frequency separation in Fig. 1b-d. As the rod-end gap increases, the frequency spread of the mode mixing increases, as depicted by the broadening of the dip in form factor, and the increased separation in frequency between the modes at the crossing.[16] When the gap is zero, there is no mode mixing as seen in Fig. 1a by an absence of a dip in the form factor at the mode crossing.

Counterintuitively, the quality factor of the modes does not decrease with the increase in gap size, indicating the mode bandwidths do not increase to account for the frequency spread. Instead, the larger rod-end gap increases the electric potential due to the capacitance effect, causing the mixing to start to occur at a great frequency difference between the two modes.

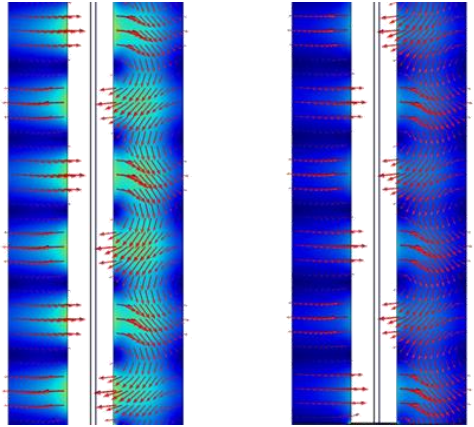

**Fig. 2** Cross-section of electric field of the mixed modes during mode crossing. Two modes are shown with the center white area of each being the location of the tuning rod (no field). The blue area is the field strength, with lighter color indicating stronger field. The cavity wall and endcaps would touch the outer edge of the blue areas. The red arrows show the electric field vectors. The left side of both modes demonstrates a TE-like field, while the right side of both modes contain some $z$ component of the electric field. The form factor of both modes is approximately the same.

Figure 2 shows a cross-section of the electric field in the mixed modes during the mode crossing. The field strength is indicated by the blue area with a stronger field corresponding to lighter color. The white area indicates the tuning rod or cavity boundaries where there is no field. The red arrows depict the field vectors. Orthogonality breaking is observed. Both modes have some electric field in the $z$-axis, and thus neither is a TE mode. Maxwell's equations can be used to show neither mode is TM, thus they are both mixed modes. The form factors of the modes are approximately the same.

The frequency separation at the mode crossing causes a gap in the frequency scan range of the cavity. At the rod orientation where the form factor is lowest in a mode crossing, the form factors of the mixed modes are the same. At that point, a search must move from scanning one mode to scanning the other, resulting in a gap in frequency. As the separation increases, the frequency gap increases. Figure 3 shows the gap in frequency scan range, $\Delta f$, normalized by the mode frequency, as a function of rod-end gap. At small gap sizes, $\Delta f/f \approx g/L$.

Longitudinal symmetry breaking induces additional modes from degeneracy breaking and mode localization. The simulation showed reentrant modes appear in the rod-to-endcap gaps. The mode crowding interferes with tracking a search mode. Since the form factor of a mode is not directly measurable, the signal from the search mode is indistinguishable from signals from other modes with nearly the same frequency, complicating searches with haloscope detectors. Additionally, mode crowding increases mode crossings and mode mixing, further degrading the capability of a detector.

Figure 4 shows the nearest 10 modes to the lowest search mode with the tuning rod along the center axis for various rod-end gaps, depicting the mode crowding as the modes move closer together with increased gap size. The modes cluster in two distinct groups. Several new modes are observed as the gap increases (i.e., the highest frequency mode shown at $g/L = 0.007$ is a reentrant mode not observed at smaller gap sizes; two modes shown in the upper cluster at $g/L = 0.02$ are degeneracies not observed at smaller gap sizes).

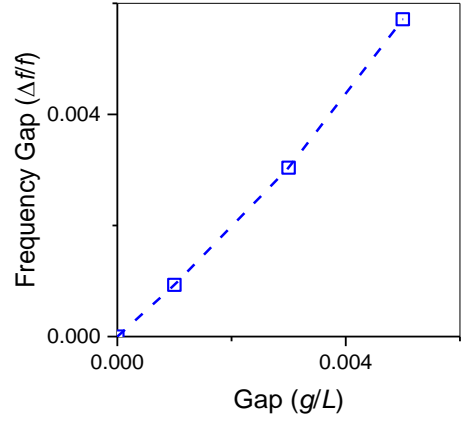

**Fig. 3** Gap in frequency scan range due to mechanical gap between the rod-ends and the cavity endcaps. The longitudinal symmetry breaking breaks the mode orthogonality, causing modes to mix when they are nearly the same frequency. An axion search must transition from one mode to another at a mode crossing, resulting in a gap in the frequency scan range. For a single tuning rod of radius $r/R \approx 0.25$, $\Delta f/f \approx g/L$ at small gaps.

Tilting of the tuning rod is a more complex symmetry breaking, as it breaks both transverse and longitudinal symmetry simultaneously. Mode localization can take on several forms and has been observed to produce degeneracy breaking in modes that were not degenerate prior to the symmetry breaking. Figure 5a-d shows the frequency of the lowest search mode and the corresponding form factor as the tuning rod is moved from the center of the cavity for tuning rod tilt ($\varphi$) of 0.25°, 0.50°, 1.00°, and 1.80°, respectively. The distance from the center of the tuning rod to the center of the cavity, $x$, is shown on the $x$-axis, and normalized by the radius of the cavity, $R$. Note, Fig. 1a shows the results for zero tilt.

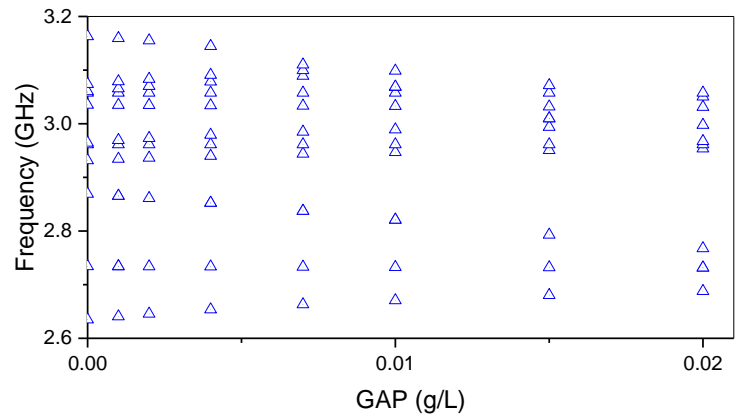

**Fig. 4** Mode crowding induced by mechanical gaps between rod-ends and endcaps. The tuning rod is located along the center axis. As the gap is increased, modes move closer together and more modes are observed due to localization and degeneracy breaking. The plot shows two distinct clustering of modes; the search mode is found in the higher-frequency cluster.

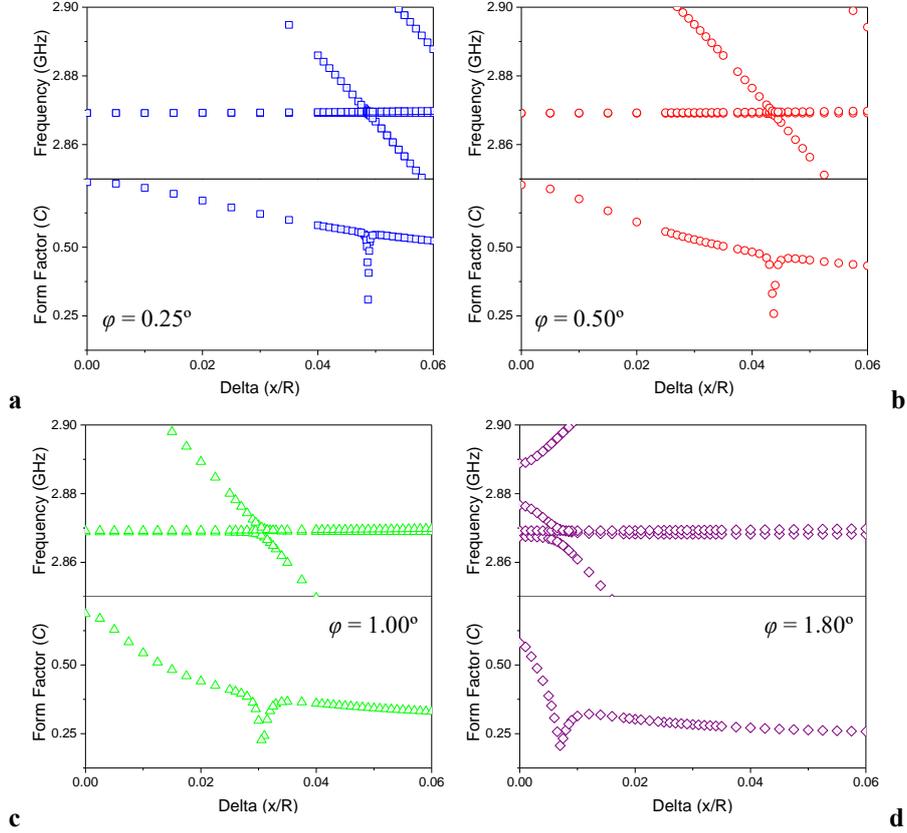

**Fig. 5** Frequency of the lowest search mode and the corresponding form factor as the tuning rod is moved through a mode crossing of the cavity for a tilt of **a)** $\varphi = 0.25°$; **b)** $\varphi = 0.50°$; **c)** $\varphi = 1.00°$; **d)** $\varphi = 1.80°$. The distance from the center of the tuning rod to the center of the cavity, $x$, is shown on the $x$-axis, and normalized by the radius of the cavity, $R$. As the tilt increases, more mode mixing is observed and the form factor is further lowered across the scan range. The next higher TM-like (non-searchable) mode is also shown for completeness.

The plots demonstrate the effects of a mode crossing and shows a degeneracy TE mode and the next higher TM-like mode. Mode mixing is depicted as a reduction in form factor and a frequency separation in modes at the crossing. Similar phenomena are observed as with the rod-end gaps. The mode mixing occurs within a smaller frequency span than from rod-end gaps, but the reduction in form factor across the scan range is more significant. Mode crowding was also observed, though less severe than with rod-end gaps.

At a tilt above ~1.00°, higher-order tuning modes obtain a greater form factor than the lowest tuning mode during part of the scan range. At a tilt of 1.80°, the next higher TM-like mode increased frequency initially and then obtains a higher form factor ($C \approx 0.3$) approximately when the lowest tuned mode begins to mix at the mode crossing. This effect produces several large gaps in the scan range.

## 4 Conclusion

The numerical analysis presented here showed orthogonality breaking in microwave cavity modes is a result of longitudinal symmetry breaking and is the cause of mode mixing. As the symmetry breaking increases, mode mixing, mode crowding, and the gap in frequency scan range increase, reducing the effectiveness of a haloscope detector. For the geometry analyzed, the ratio of the frequency gap to the mode frequency ($\Delta f/f$) was approximately equal to the ratio of the rod-end gap to the cavity height ($g/L$) for small gaps. Rod tilting caused complex mode mixing. At a tilt above ~1.00°, the most efficient search mode changed depending on rod orientation and multiple gaps in frequency scan range are observed.

**Acknowledgements** This research was supported by DOE grant DE-SC0010296.